\documentclass[twocolumn,showpacs,aps,floatfix]{revtex4}
\usepackage{amsmath}
\usepackage{graphicx}% Include figure files
\usepackage{dcolumn}% Align table columns on decimal point
\usepackage{bm}% bold math

\begin{document}

\title{Detection of  NMR signals with a radio-frequency atomic magnetometer}
\author{I. M. Savukov, S. J. Seltzer and M. V. Romalis}
\affiliation{Department of Physics, Princeton University, Princeton,
New Jersey 08544}

\date{\today}

\begin{abstract}
We demonstrate detection of proton NMR signals with a radio
frequency atomic magnetometer tuned to the NMR frequency of 62 kHz.
High-frequency operation of the atomic magnetometer makes it
relatively insensitive to ambient magnetic field noise. We obtain
magnetic field sensitivity of 7 fT/Hz$^{1/2}$ using only a thin
aluminum shield. We also derive an expression for the fundamental
sensitivity limit of a surface inductive pick-up coil as a function
of frequency and find that an atomic rf magnetometer is
intrinsically more sensitive than a coil of comparable size for
frequencies below about 50 MHz.

\end{abstract}
 \maketitle
\section{Introduction}

Nuclear magnetic resonance (NMR) signals are commonly detected with
inductive radio-frequency (rf) pick-up coils. Recently, alternative
detection methods using SQUID magnetometers
\cite{Pines,Bussell,Greenberg} or atomic magnetometers
\cite{BudkerNMR,NMRPRL} have been explored. These techniques can
achieve higher sensitivity at low NMR frequencies and offer other
advantages in specific applications. In particular, atomic
magnetometers eliminate the need for cryogenic cooling and allow
simple multi-channel measurements~\cite{RomalisNature}. However,
most atomic magnetometers are designed to detect quasi-static
magnetic fields and are sensitive to oscillating fields only in a
limited frequency range. Previous NMR and MRI experiments with
atomic magnetometers detected either static nuclear magnetization
\cite{Cates,BudkerNMR,BudkerMRI} or nuclear precession at a very low
frequency ($\sim$ 20 Hz) \cite{NMRPRL}.

Recently we developed an rf atomic magnetometer that can be tuned to
detect magnetic fields at any frequency in the kHz to MHz range
\cite{HighFreqPRL} and demonstrated detection of NQR signals at 423
kHz using this device \cite{NQR}. Another technique for detection of
rf fields with atoms is presented in \cite{BerkeleyRF}. Here we
describe detection of NMR signals from water at 62 kHz and discuss
issues specific to NMR detection, such as application of a uniform
static magnetic field. The rf magnetometer offers a number of
advantages over traditional quasi-static atomic magnetometers. It
can detect NMR signals in a wide range of magnetic fields and allows
measurements of chemical shifts~\cite{SQUIDchemshift}. Operation at
high frequency reduces the magnetic noise produced by Johnson
electrical currents in nearby conductors \cite{Varpula}. In magnetic
resonance imaging applications it increases the available bandwidth
and eliminates the effects of transverse magnetic field gradients
\cite{PinesMRI}. The rf magnetometer also has a number of practical
advantages.  It is relatively insensitive to changes in DC magnetic
field allowing it to operate in an unshielded or lightly shielded
environment. Unlike previous setups, we did not use $\mu$-metal
magnetic shields in this experiment, relying only on a thin aluminum
rf shield. The magnetometer is also relatively insensitive to
vibrations and laser noise because it detects alkali-metal spin
precession signals at high frequency. We used inexpensive multi-mode
diode lasers mounted on an aluminum plate without vibration
isolation. We identified several technical issues that need further
research, such as improvement in the uniformity of the static
magnetic field and reduction of the magnetometer dead time after the
rf excitation pulse.

We also compare the fundamental limits on the magnetic field
sensitivity for an rf magnetometer and a traditional inductive
pick-up coil. We derive an estimate for the sensitivity of a surface
pick-up coil over a wide frequency range and compare its optimal
performance with that of an atomic rf magnetometer of similar size.
We find that the fundamental sensitivity of an atomic magnetometer
is higher than fundamental sensitivity of a pick-up coil for
frequencies below about 50 MHz.

\begin{figure}
%[tbp]
\centerline{\includegraphics*[scale=0.6]{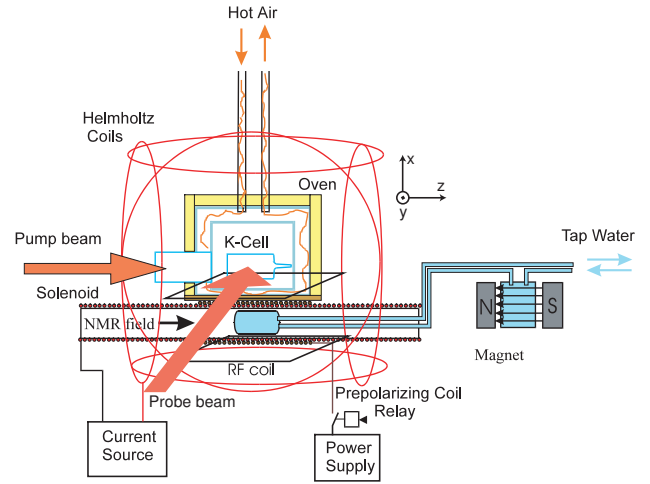}}
\caption{Experimental setup for the observation of water NMR with a
radio-frequency atomic magnetometer.}
 \label{aparatus}
\end{figure}

The principle of operation of the rf alkali-metal magnetometer is
discussed in \cite{HighFreqPRL}. Briefly, it uses a bias magnetic
field to tune the Zeeman resonance frequency of alkali atoms
$\nu=\gamma B$ (potassium with nuclear spin $I=3/2$ has
$\gamma=g\mu_B/\hbar (2I+1)=700$ kHz/Gauss) to the frequency of the
oscillating magnetic field.  The alkali atoms are optically pumped
along the bias field and their transverse spin precession excited by
the weak rf field is detected with an orthogonal probe laser. The
experimental setup for the magnetometer is shown in Figure 1.
Helmholtz coils are used to cancel the Earth field and generate the
bias field. The alkali metal is contained in a glass cell that is
heated to about 180$^{\circ}$C with flowing hot air. The sensitivity
of the magnetometer near its resonance frequency is shown in Figure
2. The broad peak in the spectral density of the magnetometer signal
is due to transverse spin oscillations excited by magnetic field
noise and other sources of fluctuations. The width of the peak is
equal to the bandwidth of the magnetometer and its height indicates
the level of magnetic noise. In Figure 2 we compare the noise levels
of the rf magnetometer operating in an unshielded environment, with
simple eddy-current shielding using thin aluminum sheets, and inside
multi-layer magnetic mu-metal shields. For comparison, with a
quasi-static magnetometer operating in an unshielded environment we
observed noise of several pT/Hz$^{1/2}$ \cite{OpenMagn}. The
degradation of the performance for an unshielded rf magnetometer is
much smaller than for a quasi-static magnetometer and the noise can
be further reduced using a more rf-tight aluminum box.

\begin{figure}
%[tbp]
\centerline{\includegraphics*[scale=1]{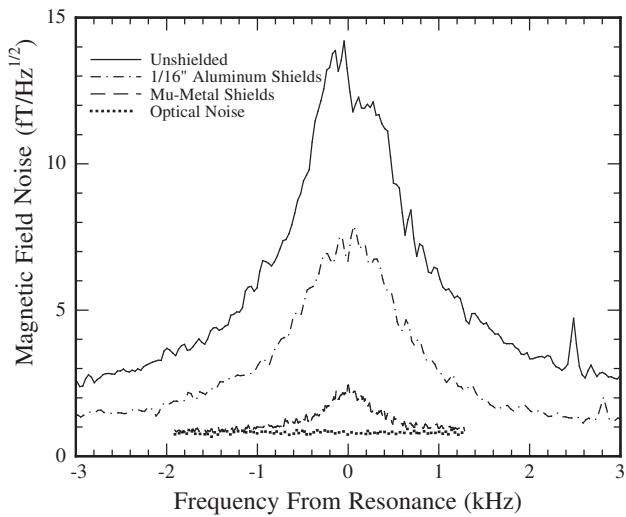}}
\caption{Comparison of sensitivities of high-frequency atomic
magnetometer: 1) unshielded (solid line), at 75 kHz; 2) with 1/16"
aluminum shield (dash-dotted line), at 75 kHz; 3)  mu-metal shielded
(dashed line), at 99 kHz; 4) optical noise (dotted line). The NMR
signals were detected using the aluminum shield. }
 \label{compnoise}
\end{figure}

For NMR detection the water sample was contained in a 3 cm diameter
and 4 cm long cylindrical glass cell. The cell was placed in a
solenoid which created a large field inside for the nuclear spins
while producing relatively little field outside~\cite{BudkerNMR,
NMRPRL}. The solenoid is needed to match the resonance frequency of
the nuclear spins to the Zeeman resonance of the atomic
magnetometer. For efficient detection of the NMR signal the diameter
of the solenoid should be close to the diameter of the water sample.
At our NMR frequency of 62 kHz the solenoid also needs to have a
relatively high magnetic field homogeneity, on the order of
$10^{-5}$, to obtain free induction decay time $T_2^*$ for water
close to the intrinsic transverse spin relaxation time. It is
relatively easy to make the solenoid long enough or add
end-correction coils so that the ends of the solenoid have a
negligible effect on the field homogeneity.  But we found that near
the center of the solenoid the field non-uniformity is limited by
variations in the pitch of the winding. Several solenoids were wound
on a 3.5 cm OD G10 tube with standard gauge AWG22 magnet wire using
different methods, including a lathe with automatic feed control.
However, after several winding attempts we were not able to obtain
longitudinal field homogeneity better than $1\times 10^{-3}$ over a
3 cm region. The field variation along the axis of the solenoid had
a rather random pattern. We suspect the non-uniformity is caused by
imperfections in the shape of the wire or non-uniform thickness of
enamel insulation. Only a 0.5 $\mu$m  variation in the winding pitch
is sufficient to explain observed non-uniformity. To improve the
field homogeneity we added a short shimming solenoid around the cell
which had a variable wire pitch. With appropriate current in the
shimming solenoid the field homogeneity along the axis was improved
to $1\times 10^{-4}$. However, the improvement in the NMR linewidth
was smaller because the shimming solenoid generated large field
gradients away from the solenoid axis. The maximum $T_2^*$ for water
NMR signals we were able to obtain was equal to 9 ms at 62 kHz. More
work will be needed in the future on the development of magnetic
field coils that create highly uniform fields over a large fraction
of their volume.

We also investigated in detail the dead time of the atomic
magnetometer following an rf pulse needed to tip the nuclear spins.
For a simple linear system the dead time is closely related to the
bandwidth of the response. As shown in Figure 2, the bandwidth of
the atomic magnetometer is on the order of 1 kHz and hence one would
expect a dead time on the order of 1 ms. However, for large
excitations the response of the rf atomic magnetometer is
non-linear. The decay time of the transverse atomic spin
oscillations becomes shorter when the longitudinal spin polarization
is reduced due to the effect of fast spin-exchange
collisions~\cite{HighFreqPRL}. Also, because the atomic vapor is
optically thick, the propagation of the pumping laser is affected by
the degree of atomic spin polarization. If the atoms are completely
depolarized during the rf pulse, the time it takes to repump the
vapor is proportional to the number of atoms and inversely
proportional to the photon flux~\cite{EffPRL}. In our conditions  it
takes about 10~ms for the pump laser to polarize the atoms and
propagate through the cell.

There are several methods for reducing the dead time of the
magnetometer. One can construct a special rf coil
\cite{onesidedRF,NQR} that creates a large magnetic field for the
nuclear spins while generating only a small field at the location of
the atomic magnetometer. One can temporarily change the bias
magnetic field experienced by the atomic magnetometer so the rf
pulse is no longer on resonance for the atomic spins and does not
cause significant spin excitation \cite{NQR}. One can also increase
the power of the  pumping laser to reduce the transverse spin
relaxation time - similar to Q-damping techniques used with
traditional rf pick-up coils. We have explored the last two
techniques here. We verified that the alkali-metal polarization is
preserved if the magnetic field is detuned sufficiently far during
the rf pulse. It is important that the bias magnetic field is only
changed in magnitude, but not in direction, in order not to excite
transverse atomic spin oscillations. It is also important to
minimize the amount of conductive materials in the vicinity of the
magnetometer to reduce eddy currents that prevent quick changes in
the bias magnetic field, but this was not easy in our setup. We
found that in the existing setup the simplest way to reduce the dead
time was by increasing the intensity of the pump laser and reducing
the density of alkali-metal atoms. This reduced the repumping time
after the rf pulse to about 3~ms but at the same time also reduced
the sensitivity of the magnetometer.

To further reduce the effect of dead time, NMR signals were acquired
using a spin-echo sequence: a $\pi/2$ pulse followed by a $\pi$
pulse after a time $\tau=15$ msec. The NMR signals were collected in
two modes, either using water that was pre-polarized by flowing it
through a permanent magnet with a field of 140 mT or using water
that was polarized in-situ by a brief application of a 10 mT
magnetic field created by a separate pre-polarizing coil wound
around the sample. For the flow-through mode the turbulent water
motion resulted in an increase of the effective diffusion constant,
so successive spin-echo signals had an effective $T_2$ decay
constant of 140 msec, which resulted in a slight decrease of the
signal.

The atomic rf magnetometer is intrinsically sensitive to a magnetic
field rotating in the same direction as the alkali-metal atoms
\cite{HighFreqPRL}. Nuclear spin precession generates dipolar fields
that can be generally decomposed into two counter-rotating
components of unequal magnitudes. To obtain the largest NMR signal,
the direction of the magnetic field in the solenoid is chosen  so
the larger of the two rotating NMR field components is co-rotating
with the atomic spins.

\begin{figure}
%[tbp]
\centerline{\includegraphics*[scale=0.8]{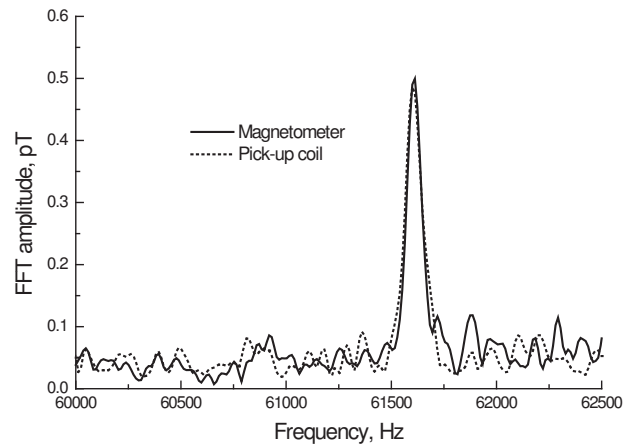}}
\caption{Comparison of the water NMR signal obtained after 10
averages with the rf atomic magnetometer (solid line) and a
traditional inductive pick-up coil (dashed line). The water is
prepolarized by flow through a 0.14T permanent magnet. Both the
atomic magnetometer and the NMR coil are located 5 cm away from the
water sample.  The active volume of the atomic magnetometer is equal
to 0.5 cm$^3$, while the volume of the pick-up coil winding is 19
cm$^3$.} \label{fftflow}
\end{figure}

In Figure 3 we show the Fourier transform of the NMR signal detected
with the atomic magnetometer compared with the signal detected with
a traditional rf pick-up coil, each after 10 averages. In Figure 4
we show the time-domain spin-echo signal obtained with the rf atomic
magnetometer after 100 averages. Finally, in Figure 5 we show the
NMR signal detected from water that is pre-polarized in-situ by
application of a 10 mT magnetic field for 2.5 sec before each
spin-echo pulse.

\begin{figure}
\centerline{\includegraphics*[scale=0.4]{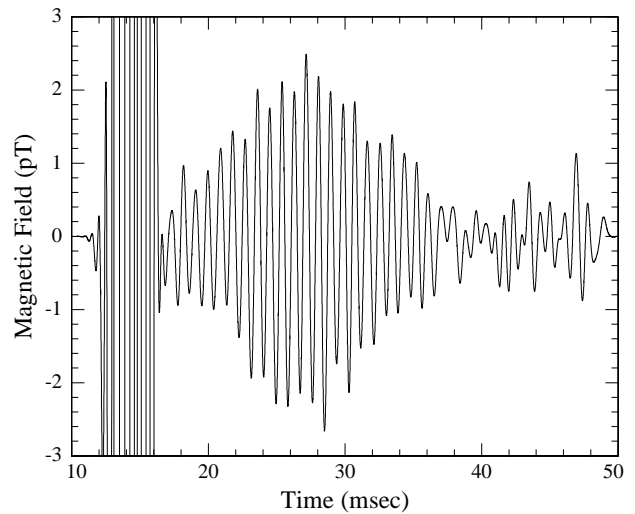}}
\caption{NMR signal from water pre-polarized by flow through an
external 0.14 T permanent magnet after 100 averages. The NMR signal
is mixed with a reference signal at 60.5 kHz and filtered with
bandwidth of 2 kHz.  The $\pi/2$ pulse is 0.5 msec long and is
applied at $t=0$, the $\pi$ pulse is applied at 12 msec. The
recovery time of the magnetometer after the rf pulse is about 3
msec. }
 \label{timedomain}
\end{figure}

\begin{figure}
%[tbp]
\centerline{\includegraphics*[scale=0.8]{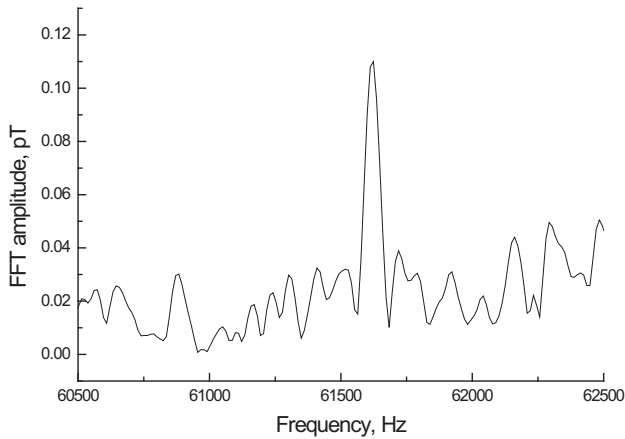}}
\caption{FFT of NMR signal after 50 averages from water
pre-polarized in-situ by application of a 10 mT magnetic field
generated by a small solenoid wound around the water sample. The
noise level of the magnetometer under these conditions is 20
fT/Hz$^{1/2}$.}
 \label{prepolfft}
\end{figure}

It can be seen in Fig. 3 that in our data the signal-to-noise ratio
obtained with the atomic magnetometer is comparable to that obtained
with a much simpler inductive pick-up coil.  Therefore, one can ask
under what conditions is the atomic magnetometer advantageous for
detection of NMR? To answer this question, we compare the
fundamental sensitivity limits for an atomic magnetometer and an
inductive  pick-up coil. The fundamental sensitivity limit for an
atomic magnetometer is derived in \cite{HighFreqPRL}. It is given by
\begin{equation}\label{ult}
\delta B_{at}=\frac{2}{\gamma}\sqrt{\frac{ \bar{v}
[\sigma_{ex}\sigma_{sd}/5]^{1/2}}{V_a}(1+\frac{1}{4\sqrt{\eta}})},
\end{equation}
where $\gamma$ is the atomic gyromagnetic factor, $\bar{v}$ is the
average thermal velocity, $\sigma_{ex}$ and $\sigma_{sd}$ are the
spin-exchange and spin-destruction collision cross sections for
alkali atoms, $V_a$ is the active volume of the atomic magnetometer
and $\eta$ is the quantum efficiency of the photodetectors. The
numerical coefficients in Eq.(\ref{ult}) are specific to alkali
atoms with nuclear spin $I=3/2$ such as K.  In our setup the active
volume of the magnetometer is equal to 0.5 cm$^3$ and the
fundamental noise limit given by Eq.~(\ref{ult}) is equal to 0.14
fT/Hz$^{1/2}$, using relaxation cross-sections for K atoms given
in~\cite{HighFreqPRL}.

\begin{figure}
\centerline{\includegraphics*[scale=0.5]{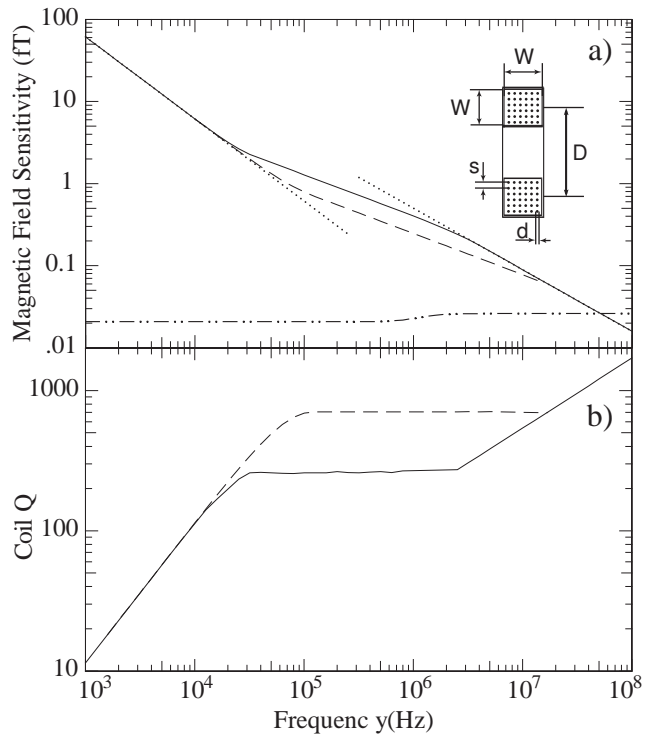}}
\caption{Panel a) Estimated optimal magnetic field sensitivity for a
surface pick-up coil with dimensions $D=5$ cm and $W=1$ cm. Solid
curve gives the sensitivity for solid wire, while dashed curve is
for Litz wire with 1000 strands. The total number of turns and the
diameter of the wire is optimized at each frequency. Dotted lines
show the asymptotic sensitivity of the coil at low frequency from
Eq. (\ref{lowf}) and high frequency from Eq. (\ref{highf}).
Dot-dashed line shows sensitivity for a K atomic magnetometer
occupying the same volume as the coil. Panel b) The $Q$ of the coil
with parameters that give optimal sensitivity in panel a). Solid
line is for solid wire, dashed line for Litz wire} \label{coilsens}
\end{figure}

Here we derive a relationship for the magnetic field sensitivity of
a surface pick-up coil valid in kHz to MHz frequency range. To our
knowledge, such general relationship has not been reported
previously in the literature. We focus on the surface coil
arrangement since for applications using such geometry the coil can
be replaced directly with an atomic magnetometer cell. We consider a
coil with a mean diameter $D$ and a square winding cross-section of
size $W\times W$ with $W \ll D$, as shown in the inset of
Fig.~\ref{coilsens}. First we consider the low frequency limit,
where eddy current losses and parasitic capacitance between coil
turns can be neglected. Suppose the coil contains $N$ turns of wire
with diameter $d$ that fill the available winding volume $V_w = \pi
D W^2$. We ignore small effects due to imperfect filling of the
winding volume by circular wire so the number of turns in the coil
is $N=4 W^2/\pi d^2$. The voltage induced in the coil by a uniform
magnetic field oscillating at frequency $\omega$ is given by $V=B
\omega \pi D^2 N/4 $, while the Johnson noise spectral density is
given by $V_n =\sqrt{16 k_B T \rho D N/d^2}$, where $\rho$ is the
resistivity of the wire material, $k_B$ is the Boltzman constant and
$T$ is the temperature. Combining these relations we get magnetic
field sensitivity limited by Johnson noise
\begin{equation}
\delta B_l=\frac{8}{\omega D} \sqrt{\frac{k_B T \rho}{V_w}}.
\label{lowf}
\end{equation}
Thus, ideal sensitivity of an inductive coil scales with the winding
volume $V_w^{1/2}$, similar to an atomic magnetometer, as well as
the frequency and the diameter of the coil. The low frequency limit
breaks down for frequencies above a few tens of kHz due to eddy
current loses, but Eq. (\ref{lowf}) is useful in giving the best
possible sensitivity for a pick-up coil.

At higher frequencies we must consider the effects of eddy currents
as well as the parasitic capacitance between coil turns. We follow a
detailed treatment of the skin depth and proximity effects for
circular wires given by Butterworth \cite{Butterworth}. For a
multi-layer coil the total effective AC resistance can be written as
\begin{equation}
R_{ac}=R_{dc} \left(1+F(z)+ u(N) \frac{d^2}{s^2} G(z)\right),
\label{AC}
\end{equation}
where the next to last and the last terms describe the skin-depth
and the proximity effects respectively. The proximity effect is
generally larger than the skin-depth effect for multi-turn coils.

The functions
 $F(z)$ and $G(z)$ are given by the ratio of Bessel functions
\begin{eqnarray}
F(z)&=& -\frac{z^2}{8} {\rm Im} \frac{ J_3(z \sqrt{- i} )}{J_1(z \sqrt{- i} )},  \nonumber \\
G(z)&=& -\frac{z^2}{8} {\rm Im} \frac{ J_2(z \sqrt{- i} )}{ J_0(z
\sqrt{- i})}.
\end{eqnarray}
Here $z= d/(\delta \sqrt{2})$, $\delta = (2 \rho/\omega
\mu_0)^{1/2}$ is the skin depth of the rf field in the conductor,
and $s$ is the spacing between the centers of the wires, $s\geq d$.
The function $F(z)$ and $G(z)$ are small for $z<1$ and grow linearly
for $z>1$ with the following asymptotic expansions:
\begin{eqnarray}
F(z)&=&z^4/192  \,\,\,\,\,\,\, {\rm for}\,\, z<1, \nonumber \\
G(z)&=&z^4/64 \,\,\,\,\,\,\, {\rm for} \,\, z<1, \nonumber \\
F(z)&=&(\sqrt{2} z - 3)/4 \,\,\,\,\,\,\,  {\rm for} \,\,z>3, \nonumber \\
G(z)&= &(\sqrt{2} z - 1)/8 \,\,\,\,\,\,\, {\rm for} \,\, z>3.
\end{eqnarray}

The function $u(N)$ depends on the winding cross-section and is
given by
\begin{eqnarray} u(N)&=&\frac{1}{N} \sum_{i=1}^{N}\left[ \left(
\sum_{j=1,j \neq i}^{N} \frac{x_j-x_i}{(x_j-x_i)^2+(y_j-y_i)^2}
\right)^2 \right.  \nonumber \\  &+& \left. \left( \sum_{j=1,j \neq
i}^{N} \frac{y_j-y_i}{(x_j-x_i)^2+(y_j-y_i)^2} \right)^2 \right],
\end{eqnarray}
where $x_i$ and $y_i$ are the positions of the wires in the
cross-section of the winding, measured in units of $s$. For a square
winding cross-section with uniform wire spacing $u(N)\approx 1.5 N$
for $N \gtrsim 20$, while for a circular cross-section $u(N)
\rightarrow (\pi/2) N$ for large $N$. For single layer coils, either
in the shape of a short solenoid or a flat spiral coil, $u(N)=3.2$
for large $N$.

Since Eq. (\ref{lowf}) does not depend on the thickness of the wire,
it seems possible to reduce skin effect and proximity effect losses
by using a very thin wire and a large number of turns to fill the
winding volume $V_w$. However, another limitation comes from
parasitic capacitance effects. For a multi-layer coil the parasitic
capacitance is dominated by the capacitance between layers. We model
it by assuming that the coil can be separated into $N_l$ winding
layers with self-inductance $L_i$, resistance $R_i$ and parasitic
capacitance $C_i$ in parallel with each layer \cite{gabriele}. Each
layer has $N_w$ wires and $N=N_w N_l$. For a surface coil geometry
with $W\ll D$ the mutual inductance between layers is approximately
equal to their self-inductance, $M_{ij}=L_i$. For a current $I$
flowing through the coil, the flux through each element $\Phi_i
=\sum_{j \neq i} M_{ij} I + L_i I = N_l L_i I$. Then the total
impendence of the coil is $Z = N_l (R_i+i \omega N_l L_i) /(1+
(R_i+i \omega N_l L_i) i \omega C_i)$. Hence, the coil has  a
self-resonance frequency given by $\omega_{self}=1/\sqrt{N_l L_i
C_i}$.

For a surface coil with $W \ll D$ the self-inductance of each layer
is $L_i=\mu_0 D N_w^2 p(D/W)$, where $p(D/W)$ is a slowly varying
dimentionless function on the order of unity. The capacitance
between layers is approximately given by $C_i = \epsilon_0 \epsilon
\pi D W /s $, where $s =W/N_l$ is the distance between layers and
$\epsilon$ is the relative permeability of the insulation material
between wires. Combining these relationships we find the
self-resonance frequency $\omega_{self}= c/(D N \sqrt{\pi \epsilon
p(D/w)})$, where $c$ is the speed of light. Thus, up to factors of
order unity, the self-resonance frequency is determined by the
duration of current propagation in the total length of the wire,
similar to other types of electromagnetic resonators. In real coils
the parasitic capacitance and mutual inductance vary between turns,
resulting in significant broadening of the self-resonance. To obtain
a high $Q$, coils are always operated significantly below their
self-resonance frequency and the number of turns is limited, $N\ll
c/D \omega$.

Another common technique for improving coil performance is to use
Litz wire made of many strands of very thin wire connected in
parallel. This reduces skin effect and proximity effect losses while
keeping the total number of turns small. Eddy current losses in Litz
wire were also considered by Butterworth in \cite{Butterworth}.
Eq.~(\ref{AC}) is slightly modified,
\begin{equation}
R_{ac}=R_{dc} \left(1+F(z)+ (u(N)+2) \frac{n^2 d^2}{s^2} G(z)\right)
\label{AC1}
\end{equation}
where $n$ is the number of strands in the Litz wire and $d$ is the
diameter of each strand, also used for evaluation of $z$. Litz wire
is effective for frequencies below a few MHz, at higher frequencies
it is not practical to make wire with $d<\delta$.

At higher frequencies one is forced to operate in the regime
$d>\delta$ and a simplified equation for the magnetic field
sensitivity can be derived using the asymptotic dependence of $F(z)$
and $G(z)$ functions for large $z$,
\begin{equation}
\delta B_h=\frac{8}{\omega D} \sqrt{\frac{k_B T \rho}{1.8 \pi D W
\delta}}. \label{highf}
\end{equation}
This can be intuitively understood as the modification of Eq.
(\ref{lowf}) for the case when the current flows only in the surface
of the coil winding within a skin depth $\delta$.

To estimate the optimal performance for a surface coil we varied the
number of turns and the wire diameter for given values of $D$ and a
square winding cross-section $W \times W$. We find that other
cross-sections, such as a single layer spiral or solenoidal coil
with the same winding width $W$ give a worse performance. For a
given application the optimal dimensions of the coil are determined
by the distance to the sample and its dimensions.  The parameters of
the coil are optimized subject to constraints that $\omega \leq 0.5
\,\omega_{self}$ and the wire diameter (including strand diameter in
Litz wire) is greater than 30~$\mu$m. We considered both solid wire
and Litz wire with 1000 strands. In Figure \ref{coilsens} we plot
the expected magnetic field sensitivity for a surface pick-up coil
with $D=5$ cm and $W=1$ cm. We also plot the $Q$ of the coil,
showing that the model predicts values of $Q$ that are consistent
with or slightly higher than common experimental values of $Q$ for
room temperature copper coils, as would be expected for an idealized
model. At high frequency $Q$ increases as $f^{1/2}$ \cite{Hoult}. It
can be seen that the magnetic field sensitivity of a surface coil is
well approximated by asymptotic relationships (\ref{lowf}) and
(\ref{highf}) for frequencies below 30 KHz and above 10 MHz
respectively. In the intermediate frequency range the sensitivity is
limited by the self-resonance effects or the minimum practical wire
diameter. As can be seen in Fig.~\ref{coilsens}, Litz wire improves
magnetic field sensitivity by about a factor of 2 in this regime.

We also plot in Figure \ref{coilsens} the expected optimized
sensitivity for an atomic magnetometer occupying the same space as
the coil with $V_a = \pi (D+W)^2 W/4$. The atomic magnetometer
sensitivity has a slight frequency dependence due to changes in
spin-relaxation when the resonance frequency becomes higher than the
spin-exchange rate, as discussed in \cite{HighFreqPRL}. We find that
a room temperature copper RF coil overtakes the sensitivity of a K
magnetometer at approximately 50 MHz.

The model for surface coil sensitivity also gives a good estimate
for the sensitivity of the pick-up coil used in our NMR experiment.
For detection of NMR signals shown in Figure 3 we used a coil with
an average diameter of 3.6 cm, cross-section of 1$\times$1.6 cm$^2$,
wire diameter $d=0.23$ mm, wire spacing $s=0.5$ mm, and $N=400$. For
these parameters, the estimated coil magnetic field sensitivity is
about 3 fT/Hz$^{1/2}$ at 66 kHz, close to the optimum for given coil
dimensions. This theoretical estimate is in a good agreement with
actual magnetic noise measurements when the pick-up coil was placed
in a well-shielded aluminum box. When the coil was used in the NMR
setup, the measured noise level was 7 fT/Hz$^{1/2}$, equal to the
noise level measured by the atomic magnetometer under the same
shielding conditions. Thus, it is not surprising that the noise
level in Figure 3 is the same for the atomic magnetometer and the
pick-up coil, both being limited by external noise sources. With
better eddy-current shielding or by using a gradiometric measurement
\cite{RomalisNature} one can expect to significantly reduce the
noise of the atomic magnetometer while the pick-up coil is already
operating near the fundamental limit of its sensitivity.

It is also interesting to compare the sensitivity of an rf atomic
magnetometer with that of a SQUID magnetometer. While at low
frequencies SQUID detectors typically have sensitivity of about
1~fT/Hz$^{1/2}$, at frequencies above a few kHz a tuned
superconducting resonator can be used to improve their performance
~\cite{Bussell,Bussel2}. For example, magnetic field sensitivity of
0.035 fT/Hz$^{1/2}$ has been demonstrated at 425 kHz using a 5-cm
diameter superconducting pick-up coil and a resonator with $Q=10^5$
\cite{Bussel2}. Atomic magnetometers have not yet experimentally
reached this level of sensitivity, although it is possible based on
fundamental sensitivity limits given by Eq. (\ref{ult}). On the
other hand, they have a larger bandwidth, which is important for
many applications. For example, in \cite{NQR} magnetic field
sensitivity of 0.24 fT/Hz$^{1/2}$ has been demonstrated for an rf
atomic magnetometer with $Q$ of $10^3$ at 423 kHz.

In conclusion, we have demonstrated the first detection of proton
NMR signals with an rf atomic magnetometer. The advantages of this
technique include relative insensitivity of the rf magnetometer to
ambient magnetic field noise and the possibility of measuring NMR
chemical shifts. We also demonstrated in-situ prepolarization of
proton spins, opening the possibility of efficient magnetic
resonance imaging with an atomic magnetometer that does not rely on
remote encoding \cite{BudkerMRI}. We identified several issues that
need further improvements, such as better magnetic field homogeneity
for compact solenoids and more efficient damping of magnetometer
spin transients. Finally, we derived a simple relationship for
estimating the magnetic field sensitivity of a surface coil over a
wide frequency range. Comparing it with that of an atomic
magnetometer we find that atomic magnetometers have an intrinsic
sensitivity advantage over a pick-up coil for frequencies below
about 50 MHz. Thus, they are well-suited for detection of NQR
signals as well as for low field NMR and MRI.

We like to thank K. Sauer for helpful discussions. This work was
supported by the NSF and the Packard Foundation.

\bibliography{highfreqnmr}

\end{document}